\documentclass[twoside]{article}
\usepackage{fleqn,espcrc2}
\usepackage[reqno]{amsmath}
\usepackage{epsf,rotate,amsfonts,amssymb}
\newcommand{\ebox}[2]{\epsfxsize=#1 \epsfbox[10 30 560 590]{#2}}

\newcommand{\Dslash}{/\hspace{-1.5ex}D}
\newcommand{\Dslashbak}{\overleftarrow{\Dslash}}

\newcommand{\pslash}{\not\!p}

\newcommand{\kslash}{\not\!k}
\newcommand{\half}{\frac{1}{2}}
\newcommand{\bra}{\langle}
\newcommand{\ket}{\rangle}
\newcommand{\Tr}{\operatorname{Tr}}

\newcommand{\order}{\mathcal{O}}

\newcommand{\khat}{\hat{k}}
\newcommand{\psibar}{\overline{\psi}}
\newcommand{\Bi}{B_{\mathit{I1}}}
\newcommand{\zz}{Z^{(0)}}
\newcommand{\dmz}{\Delta M^{(0)}}

\begin{document}

\title{The Quark Propagator in Momentum Space}

\author{Jon Ivar Skullerud and Anthony G.\ Williams\address{
CSSM, University of Adelaide, Adelaide 5000, Australia}}

\begin{abstract}
The quark propagator is calculated in the Landau gauge at
$\beta=6.0$.  A method for removing the dominant, tree-level 
lattice artefacts is presented, enabling a calculation of the
momentum-dependent dynamical quark mass.
\end{abstract}

\maketitle

\section{Introduction}

The quark propagator is one of the fundamental quantities in QCD.  By
studying the mass function, which is the scalar part
of the quark propagator, we can gain insight into the
mechanisms of chiral symmetry breaking.  The momentum dependence of
the quark propagator is also used extensively as input in
Dyson--Schwinger equations for hadronic matrix elements.  A lattice
study of the quark propagator may enable us to check the
validity of the models used in these calulations.

\section{$\order(a)$-improved quark propagator}

All $\order(a)$ errors in the fermion action can be removed by adding
terms to the Lagrangian \cite{Luscher:1996sc,Dawson:1997gp},
\begin{multline}
\mathcal{L}(x) = \mathcal{L}^W 
- \frac{i}{4}c_{sw}a\psibar\sigma_{\mu\nu}F_{\mu\nu}\psi \\
 + \frac{b_g am}{2g_0^2}\Tr(F_{\mu\nu}F_{\mu\nu})
 - b_mam^2\psibar\psi \\
 + c_1a\psibar\Dslash^2\psi
 + c_2am\psibar\Dslash\psi
\label{eq:improve-action}
\end{multline}
$m$ should here be taken to be the subtracted bare mass $m\equiv
m_0\!-\!m_c$.  The $b_g$ and $b_m$ terms correspond to a rescaling of
the coupling constant and the mass respectively.  The two last terms
can be eliminated by a field transformation \cite{Heatlie:1991kg},
\begin{alignat}{2}
\psi & \to \psi' =
 (1+b_qam)(1-c_qa\Dslash)\psi && \equiv L\psi \notag \\
\psibar & \to \psibar' =
(1+b_q am)\psibar(1+c_qa\Dslashbak) && \equiv \psibar R
\label{eq:rotate}
\end{alignat}
The tree level improved action after the
transformation (\ref{eq:rotate}) has $c_{sw}=1,
b_q=\frac{1}{4}$ and $c_q=\frac{1}{4}$.
The improved propagator is given by
\begin{equation} 
S(x,y) \equiv \bra\psi'(x)\psibar'(y)\ket
 = L(x)S_0(x,y)R(y)
\label{eq:rotprop}
\end{equation} 
where $S_0(x,y) \equiv \bra\psi(x)\psibar(y)\ket$.
Since the propagator $S_0$
is defined as the inverse of the fermion matrix $M(x) \equiv
\Dslash(x)+m_0+\order(a)$, 
we can use this to obtain another, simpler expression for the improved
propagator:
\begin{multline}
S(x,y) = \bigl(1+2(b_q+c_q)am\bigr)S_0(x,y)\\
 - 2ac_q\delta(x-y) + \order(a^2)
\label{eq:rotate-equiv}
\end{multline}
With this in mind, we define the tree-level improved propagator
$S_I(x,y)$ as
\begin{equation}
S_I(x\!-\!y) \equiv (1+ma)S_0(x\!-\!y) - \frac{a}{2}\delta(x\!-\!y)
\label{eq:improved-prop} 
\end{equation}
We will also introduce the tree-level `rotated' propagator
$S_R(x,y)$ as the special case of the improved propagator
in (\ref{eq:rotprop}),
\begin{gather}
S_R(x,y) \equiv (1+\frac{am}{2})L'(x)S_0(x,y)R'(y)
\end{gather}
where
$L' \equiv 1-a\Dslash/4$ and $R' \equiv 1+a\Dslashbak/4$.

\section{Analysis}

The momentum space quark propagator in a particular gauge is given by
$S(p) = \sum_x e^{-ipx}S(x,0)$.
We introduce the following `lattice momenta' 
$k_\mu = \frac{1}{a}\sin(p_\mu a)$ and
$\khat_\mu  = \frac{2}{a}\sin(p_{\mu}a/2)$
which differ by
\begin{equation}
a^2\Delta k^2 \equiv \khat^2 - k^2 
 = \frac{a^2}{4}\sum_{\mu}p_\mu^4 + \order(a^4) \;.
\end{equation}
In the continuum, the quark propagator has the following general form,
\begin{equation}
S(p) = \frac{1}{i\pslash A^c(p) + B^c(p)} \equiv
\frac{Z^c(p)}{i\pslash+M^c(p)} 
\end{equation}
We expect the lattice quark propagator to have a similar form, but
with $\kslash$ replacing $\pslash$:
\begin{equation}
S(p) = \frac{Z(p)}{i\kslash + M(p)}
\end{equation}

The dimensionless Wilson fermion propagator at tree level is
\begin{equation}
S_0^{(0)}(p) = \frac{-i\kslash a + m_0a + \half\khat^2 a^2}
{k^2 a^2 + \left(m_0a+\half\khat^2 a^2\right)^2} \, .
\end{equation}
The tree level `improved' propagator is given by
\begin{gather} 
S_I^{(0)}(p) = (1+m_0a)S_0^{(0)}(p)-\half
\end{gather}
If we write
\begin{equation}
\left(S_I^{(0)}(p)\right)^{-1} = 
 \frac{i\kslash a + m_0a + a\dmz(p)}{\zz(p)}
\end{equation}
we find
\begin{align}
\zz(p) &  =  \frac{k^2 a^2(1+m_0a)^2 + \Bi^2}{(1+m_0a)D}
\label{eq:z0def} \\
a\dmz(p)
 & = \frac{m_0^2 a^2 - a^4\Delta k^2 + a^4\khat^4/4}{1+m_0a}
\label{eq:deltam0def}
\end{align}
where
\begin{align}
D & = k^2 a^2 + \left(m_0a + \half\khat^2a^2\right)^2 \\
\Bi & =  m_0a+\frac{m_0^2a^2}{2}+\frac{a^4\Delta k^2}{2}
 -\frac{a^4\khat^4}{8}
\end{align}

The corresponding tree level expressions for $S_R$ are considerably
more complicated than those for $S_I$, and we will not reproduce
them here.

We are here primarily interested in studying the deviation of the
quark propagator from its tree level value.  Since QCD is
asymptotically free, we should at sufficiently high momentum
values get $S(p)\to S^{(0)}(p)$ up to logarithms.  We will attempt to
separate out the tree level behaviour by writing
\begin{equation}
  S^{-1}(pa) = \frac{ia\kslash + aM_r(pa) + a\dmz(pa)}{Z_r(pa)\zz(pa)}
\label{eq:zmdef}
\end{equation}
where $m_0$ is replaced by $m$ in the
expressions for $\zz$ and $\dmz$.  Asymptotically, we 
expect that $Z_r(pa)\to 1$ and $M_r(pa)\to m$ up to logarithms.

\section{Results}

The quark propagator is calculated at $\beta\!=\!6.0$ on a
$16^3\!\times\!48$ lattice, using the mean-field improved value
$c_{sw}\!\!=\!\!1.479$.  The configurations were fixed to Landau gauge
with an accuracy of $\theta_{\text{max}}\!=\!10^{-12}$.  At
$\kappa\!=\!0.137$, corresponding to $ma\!=\!0.0603$, we have
generated both $S_0$ and $S_R$.  At $\kappa\!=\!0.1381$, corresponding
to $ma\!=\!0.031$, only $S_0$ was generated.  $S_I$ is easily
constructed from $S_0$.  

When we calculate $Z$ and $M$ without factoring out the tree-level
behaviour, it becomes clear that both improved propagators ($S_I$ and
$S_R$) are completely dominated by the unphysical tree level behaviour
at high momenta.  In particular, $B\equiv ZM$ computed from $S_I$
becomes large and negative, approaching a value of $aB=-3$.  Only in
the infrared, below $pa\lesssim 0.8$, might we be able to extract
physically significant information.

\begin{figure}
\begin{center}
\setlength{\unitlength}{0.5cm}
\setlength{\fboxsep}{0cm}
\begin{picture}(14,7)
\put(0,0){\begin{picture}(7,7)\put(-0.9,-0.4){\ebox{4cm}{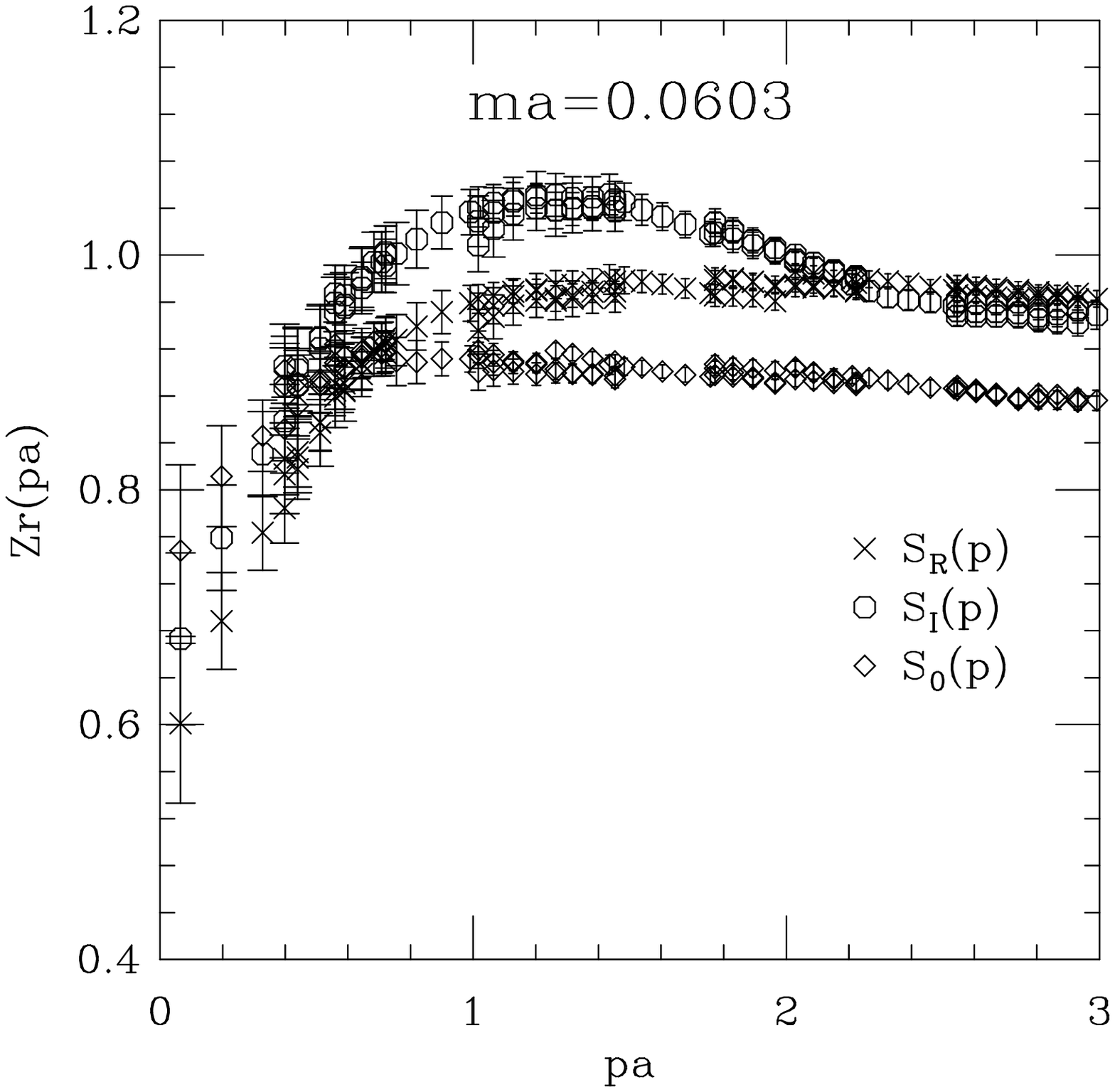}}\end{picture}}
\put(7,0){\begin{picture}(7,7)\put(-0.9,-0.4){\ebox{4cm}{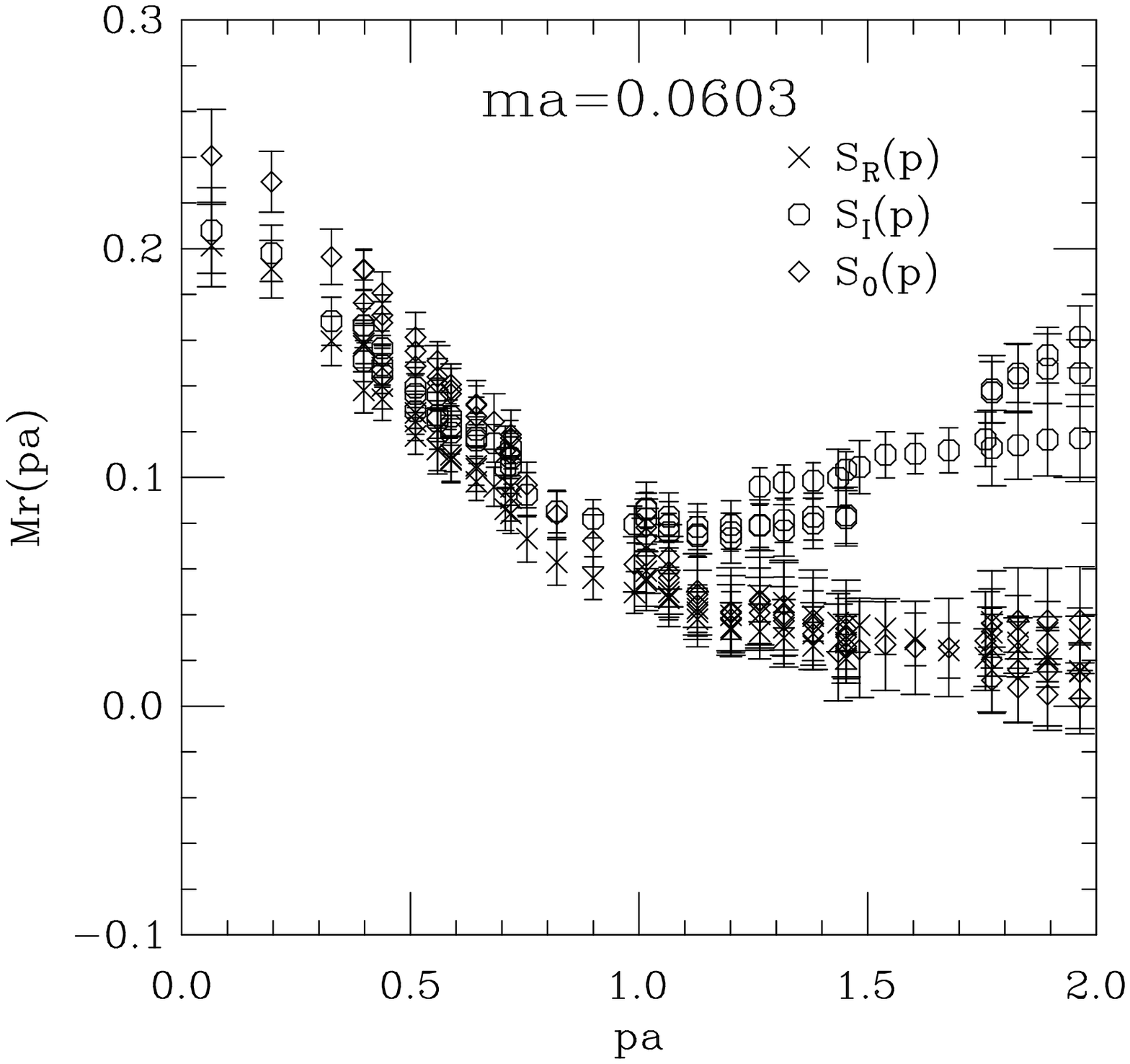}}\end{picture}}
\end{picture}
\end{center}
\vspace{-1cm}
\caption{
$Z_r(p)$ (left) and $M_r(p)$ (right) for all three propagators, from
20 configurations.
}
\label{fig:cutscompare}
\end{figure}

Fig.~\ref{fig:cutscompare} shows $Z_r$ and $M_r a$ as
functions of $pa$ for both $S_0$, $S_I$ and $S_R$.  
In order to ease the comparison and remove residual anisotropy, we
have selected momenta lying within one unit of spatial momentum of the
4-dimensional diagonal.  While all three show a dramatic improvement
on the unsubtracted data, the large momentum behaviour of the mass
function from $S_I$ is poor as a consequence of the
pathological behaviour of $S_I^{(0)}(p)$ at high momenta. This is due
to the cancellation of large terms in the subtraction for the mass.  It is
therefore desirable to use the definition $S_R$ for the improved
propagator.\footnote{A similar conclusion was reached by C.~Pittori
\cite{pittori-lat99}. }

Below $pa\sim0.9$, the values for $M_r$ agree within errors for the
two versions of the improved propagator.  In particular, the value for
the infrared mass $M_r(0)$ comes out the same.  In contrast, 
$S_0$ yields a mass which is 3--4$\sigma$ higher.

\begin{figure}
\vspace{-1cm}
\begin{center}
\ebox{7cm}{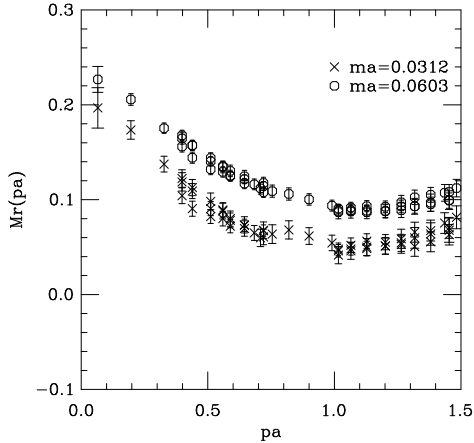}
\end{center}
\vspace{-2cm}
\caption{
$M_r(p)$ from $S_I$, for $\kappa=0.137$ (circles) and $\kappa=0.381$
(crosses).  The increase in $M_r(p)$ for $pa>1$
is an indication of the difficulty of accurately subtracting off the
tree level mass function.  The data shown are from 50 configurations.
}
\label{fig:masscompare}
\end{figure}

Fig.~\ref{fig:masscompare} shows $M_r(p)$ calculated from $S_I$ for
both quark masses.  We see that the infrared mass changes only
slightly as the bare quark mass is halved, pointing to a dynamically
generated quark mass of $(300 \pm30)$MeV in the chiral limit.
$Z_r$ turns out not to depend on the quark mass.

\section{Conclusion and further work}

We have used two different definitions of the $\order(a)$ improved
quark propagator.  We make use of asymptotic freedom to factor out the
tree level behaviour, replacing it with the `continuum' tree level
behaviour $Z(p)=1, M(p)=m$.
This tree level subtraction dramatically improves the data.  
The relatively poor behaviour of the
tree level subtracted $S_I$ can be put down to the large tree level
finite-$a$ effects which require fine tuning to subtract off
correctly.

For $pa\lesssim 1$, we see that $M_r(p)$ falls off with $p$ as
expected.  The values obtained from $S_R$ and from $S_I$ are
consistent, while those for the unimproved propagator $S_0$ differ
significantly.  We find that $M_r(0)$ approaches a value of
$300\pm30$ MeV in the chiral limit.

We also find a significant dip in the value for $Z_r(p)$ at low
momenta.  This is more pronounced for the improved propagators than
for the unimproved one.  The next step will be a quantitative study of
the functional form for $Z_r$ and $M_r$

Finite volume effects have not been studied, but there is no sign of any
anisotropy at low $p$, indicating that finite volume effects are
small.  Repeating these calculations at a different lattice spacing is
also essential especially to get reliable results for the quark mass.

{\bf Acknowledgments:} 
We thank Derek Leinweber for stimulating discussions.
Support from the Australian Research Council 
and from UKQCD Collaboration CPU
time under PPARC Grant GR/K41663 is gratefully
acknowledged

\ifx\undefined\allcaps\def\allcaps#1{#1}\fi
\providecommand{\href}[2]{#2}\begingroup\raggedright\endgroup


\begin{thebibliography}{1}

\bibitem{Luscher:1996sc}
M.~L{\"u}scher, S.~Sint, R.~Sommer, and P.~Weisz, {\em Nucl. Phys.} {\bf B478} (1996) 365,
\href{http://xxx.lanl.gov/abs/hep-lat/9605038}{{\tt hep-lat/9605038}}.

\bibitem{Dawson:1997gp}
C.~Dawson {\em et.~al.}, {\em Nucl. Phys. Proc. Suppl.} {\bf 63} (1998) 877,
\href{http://xxx.lanl.gov/abs/hep-lat/9710027}{{\tt hep-lat/9710027}}.

\bibitem{Heatlie:1991kg}
G.~Heatlie {\em et.~al.}, {\em Nucl. Phys.}  {\bf B352} (1991)
266.

\bibitem{pittori-lat99} C.~Pittori, {{\tt hep-lat/9909086}}.

\end{thebibliography}
\end{document}